\documentclass[preprint,12pt, review]{cas-sc}

\usepackage[authoryear,longnamesfirst]{natbib}
\usepackage{graphicx,hyperref}
\usepackage{float}
\usepackage{amssymb,bm,multirow,booktabs,amsmath}
\usepackage[nohyperlinks]{acronym}

\def\tsc#1{\csdef{#1}{\textsc{\lowercase{#1}}\xspace}}
\tsc{WGM}
\tsc{QE}

\begin{document}
\let\WriteBookmarks\relax
\def\floatpagepagefraction{1}
\def\textpagefraction{.001}
\RenewDocumentCommand \printorcid { } { }

\shorttitle{Geometry-adaptive Ambisonic encoding}    

\shortauthors{Xiang Zhou et~al.}  

\title [mode = title]{Geometry-adaptive Ambisonic encoding for sparse microphone arrays of variable topology using physics-informed diffusion}  



%

\author[inst1]{Xiang Zhou}[orcid=0009-0001-2270-3853]




\credit{Conceptualization, Methodology, Visualization, Validation, Writing $-$ review \& editing, Writing $-$ original draft}

\affiliation[inst1]{organization={Center of Intelligent Acoustics and Immersive Communications, School of Artificial Intelligence, Northwestern Polytechnical University}, 
            city={Xi'an},
            postcode={710072}, 
            country={China}}
\affiliation[inst2]{organization={Digital Signal Processing Lab, School of Electrical and Electronic Engineering, Nanyang Technological University},country={Singapore}}

\author[inst1]{Zhengqiao Zhao}[]
\cormark[1]
\ead{zhengqiao.zhao@nwpu.edu.cn}
\credit{Writing $-$ review \& editing, Methodology, Conceptualization}

\author[inst2]{Zhengding Luo}[orcid=0000-0002-2694-5059]

\credit{Writing $-$ review \& editing, Visualization, Conceptualization}

\author[inst1]{Wen Zhang}[orcid=0000-0002-0752-6123]


\cormark[1]

\ead{wen.zhang@nwpu.edu.cn}


\credit{Writing $-$ review \& editing, Supervision, Methodology, Funding acquisition}

\cortext[1]{Corresponding author}




\begin{abstract}
Ambisonics delivers compact scene‐based spatial‐audio representation, yet high‐order Ambisonic encoding poses difficulties for wearables and embedded hardware. Their microphone arrays are often sparse, irregular, and constrained by device-specific boundary conditions. These factors make the spherical-harmonic (SH) domain encoding ill-conditioned: inverse filtering amplifies noise, while deterministic neural encoders may overfit to array-specific responses or smooth ambiguous higher-order components. This paper presents DiffM2A, a geometry-adaptive conditional diffusion framework for robust Ambisonic encoding from sparse MAs with variable topologies. Its Geometry-Adaptive Spherical Harmonic Projection (GASHP) front-end constructs boundary-aware SH steering functions and applies an energy-normalized modal projection, mapping array-dependent observations to a
common modal representation without explicit pseudo-inverse computation. A dual-branch Elucidated Diffusion Model then estimates complex Ambisonic coefficients, conditioned on both the raw microphone spectra and GASHP features. Sound-intensity and rotational equivariance losses further enhance inter-channel phase consistency and structured behavior across SH subspaces. Evaluations on both first- and second-order Ambisonic encoding tasks, using simulated room-acoustics and real-world LOCATA recordings, demonstrate that DiffM2A outperforms conventional and neural baseline methods on signal fidelity, spectral accuracy, spatial coherence, and binaural cue preservation. Additional experiments show that these gains are largely retained across unseen five-microphone layouts and under mismatched open-array and rigid-sphere boundary models.
\end{abstract}





\maketitle


\section{Introduction}
\label{sec:intro}

Spatial audio is increasingly used in immersive media, telepresence, extended reality, and wearable devices \cite{7911385}. Ambisonics is a scene-based spatial audio format that represents a sound field using spherical harmonic (SH) components \cite{Gerzon1973PeriphonyWS,Rafaely2015FundamentalsOS}. First-order Ambisonics (FOA) is relatively easy to capture and render because it uses only a small number of channels \cite{Gerzon1975TheDO, zotter2019ambisonics}; however, its angular resolution and accurately reproduced listening region are limited. Higher-order Ambisonics (HOA) improves spatial resolution by representing the sound field with additional SH coefficients \cite{Daniel2003FurtherIO}. Nevertheless, robust acquisition of HOA remains a non-trivial challenge, as full-band encoding of supplementary SH coefficients from sparse microphone measurements is inherently ill-conditioned.

Conventional HOA recording usually depends on dense calibrated spherical arrays, whose spatial sampling matrices stay well conditioned throughout the target frequency band \cite{inproceedings06}. Wearable devices rarely satisfy these assumptions. They contain few microphones, their positions are dictated by industrial design constraints, and their geometries are often irregular \cite{10.1121/1.4795780}. When an array provides insufficient independent spatial samples for the desired Ambisonic order, the SH-domain inverse problem becomes rank-deficient or ill-conditioned. Array boundary conditions further complicate the problem. Open arrays and rigid-baffle arrays exhibit different frequency-dependent radial responses, which must be accounted for during encoding. Sparse sampling and boundary mismatch can consequently produce spatial aliasing, noise amplification, and errors in the recovered higher-order modes, particularly at mid and high frequencies \cite{4099570}.

Ambisonic encoding has been studied using both signal processing (SP) and deep learning (DL) methods \cite{Wabnitz2011UpscalingAS,Schrkhuber2019LinearlyAQ,Politis2018COMPASSCA,You2026SHBAESH,Berebi2025AmbisonicsBR,SchrkhuberBinauralRO,Gayer2025AmbisonicsEF}. Linear encoders based on least-squares estimation and pseudo-inverse filtering \cite{9909803,7952196,6645392} are effective when array transfer functions are known accurately and the sampling matrix is well conditioned. For sparse or irregular arrays, however, the inverse operation can substantially amplify sensor noise and radial-response mismatch. Regularization improves numerical stability but generally trades it for reduced spatial accuracy.
Parametric approaches replace signal-independent fixed encoding matrices with scene-dependent models of directional sound components \cite{9795678}. Although more flexible, their performance depends on reliable direction-of-arrival estimates and sparse-scene assumptions, which may be violated in reverberant or multi-source environments.

Neural encoders provide an alternative by learning a mapping from
multichannel microphone signals to Ambisonic coefficients \cite{10415885,inproceedings24am,article21,9577486,10889311,10888882}.
Representative approaches use geometry-conditioned U-Net architectures \cite{10887869}, spatial-power-map supervision \cite{10890048}, and
cross-attention conditioned on array transfer functions \cite{beyond}.
These methods reduce the need for explicit inverse filtering, but two
limitations remain. First, most existing neural encoders produce a single
deterministic estimate \cite{8643769}, although sparse microphone
observations may not uniquely determine the higher-order Ambisonic
coefficients \cite{4099570}. Under coefficient-wise regression
objectives, this ambiguity may lead to averaged or over-smoothed estimates. Second, conventional reconstruction
losses typically treat the Ambisonic coefficients independently and do not
explicitly account for spatial relationships such as low-order
sound-intensity consistency or the rotational equivariance of SH
coefficients \cite{Morgado2018SelfSupervisedGO,10890048}.

Recent studies have introduced generative models for Ambisonics upmixing and encoding \cite{wen2025guidesep, scheibler2024universal}. DiffAU uses cascaded score-based diffusion to generate higher-order channels from complete FOA signals, while SIRUP employs latent diffusion to upmix FOA steering vectors for highly directive spatialization \cite{milstein2025diffau,picard2026sirup}. Flow-HOA applies conditional flow matching to generate
deployable FIR filter banks for Ambisonic encoding from sparse microphone arrays \cite{you2026flowhoa}. These studies demonstrate the potential of generative
priors for underdetermined spatial-audio problems, but address either FOA-to-HOA upmixing or the generation of a fixed encoding filter bank. 

In contrast, we propose DiffM2A, a geometry-adaptive conditional diffusion
framework for estimating complex Ambisonic coefficients from sparse
multichannel recordings with variable array layouts. DiffM2A combines a
boundary-aware analytical projection with an observation-conditioned
diffusion estimator. Its Geometry-Adaptive Spherical Harmonic Projection
(GASHP) constructs SH steering vectors under a selected acoustic boundary
model and maps the microphone observations to a common modal representation.
Rather than explicitly inverting the array sampling matrix, GASHP applies an
energy-normalized matched-filter projection, thereby avoiding the
small-singular-value amplification associated with pseudo-inverse encoding.
The resulting features remain affected by modal scaling, cross-mode leakage,
and boundary mismatch, which are handled by a dual-branch Elucidated Diffusion
Model (EDM) conditioned jointly on the raw microphone spectra and GASHP
features. This formulation treats sparse-array Ambisonic encoding as
observation-conditioned estimation rather than a closed-form inverse problem.

The main contributions are as follows:
\begin{itemize}
  \item We introduce GASHP, a boundary-aware analytical front-end that maps
  sparse-array observations to a common SH-modal representation without
  explicit pseudo-inverse filtering.

  \item We develop a dual-branch EDM that estimates complex Ambisonic
  coefficients jointly from raw microphone spectra and GASHP features,
  providing a distribution-aware estimator for the underdetermined encoding
  problem.

  \item We introduce order-specific spatial regularization based on
  low-order sound intensity and higher-order rotational equivariance,
  encouraging inter-channel phase consistency and structured transformations
  across SH subspaces.

  \item We evaluate DiffM2A for first- and second-order Ambisonic encoding
  using simulated rooms and LOCATA recordings. Among the evaluated methods,
  DiffM2A achieves gains in signal fidelity, spatial coherence, and binaural
  cue preservation, while retaining its performance advantages on unseen
  array layouts and under mismatched open-array and rigid-sphere boundary
  models.
\end{itemize}

The remainder of this paper is organized as follows. Sec.~\ref{sec:preliminaries} introduces the SH-domain forward model and discusses the ill-conditioning of sparse Ambisonic encoding. Sec.~\ref{sec:method} presents the proposed GASHP front-end, conditional diffusion denoiser, and spatial regularization losses. Sec.~\ref{sec:exp} describes the datasets, metrics, and implementation details, followed by the experimental results and ablation analyses. Finally, the conclusion summarizes the main findings and discusses future directions.

\section{Preliminaries}
\label{sec:preliminaries}

Consider a sparse microphone array (MA) with $M$ sensors deployed in a continuous acoustic field. Let $\Omega=\{\mathbf{r}_i\}_{i=1}^{M}$ denote the array geometry, where $\mathbf{r}_i=(r_i,\theta_i,\phi_i)$ is the position of the $i$-th microphone in spherical coordinates with respect to the array origin. In the short-time Fourier transform (STFT) domain, an $N$-th order Ambisonic representation approximates the sound field as a truncated SH expansion:
\begin{equation}
p(\mathbf{r},t,f)
\approx
\sum_{n=0}^{N}\sum_{m=-n}^{n}
a_{nm}(t,f)R_n(r,f)Y_n^m(\theta,\phi),
\label{eq:sh_expansion}
\end{equation}
where $\mathbf{r} = (r, \theta, \phi)$ denotes a spatial position, $a_{nm}(t,f)$ is the Ambisonic coefficient of SH degree $n$ and order $m$, $Y_n^m(\theta,\phi)$ is a real-valued SH basis function, and $R_n(r,f)$ is a frequency-dependent radial response determined by acoustic propagation and the array boundary condition. Let $k=2\pi f/c$ denote the acoustic wavenumber,
where $c$ is the speed of sound. For an open, or free-field, array configuration, the radial response is \cite{williams1999fourier}
\begin{equation}
R_n^{\mathrm{open}}(r,f)
=
4\pi \mathrm{i}^{n} j_n(kr),
\label{eq:radial_open}
\end{equation}
where $j_n(\cdot)$ is the spherical Bessel function of the first kind. For microphones mounted on the surface of a rigid spherical baffle with radius $a$, i.e., $r_i=a$, the Neumann boundary condition gives the scattering-aware radial response \cite{williams1999fourier}:
\begin{equation}
R_n^{\mathrm{rigid}}(a,f)
=
4\pi \mathrm{i}^{n}
\left[
j_n(ka)
-
\frac{j_n^{\prime}(ka)}
{h_n^{(2)\prime}(ka)}
h_n^{(2)}(ka)
\right],
\label{eq:radial_rigid}
\end{equation}
where $h_n^{(2)}(\cdot)$ is the spherical Hankel function of the second kind, and $(\cdot)^{\prime}$ denotes differentiation with respect to the function argument. Throughout this work, HOA coefficients follow the Ambisonics Channel Numbering (ACN) convention and use Schmidt semi-normalized (SN3D) scaling \cite{nachbar2011ambix}.

Sampling the sound field in Eq.~\eqref{eq:sh_expansion} at the $i$-th microphone yields
\begin{equation}
x_i(t,f)
=
\sum_{n=0}^{N}\sum_{m=-n}^{n}
b_{i,nm}(f)a_{nm}(t,f)
+
n_i(t,f),
\label{eq:scalar_forward}
\end{equation}
where
\begin{equation}
b_{i,nm}(f)
=
R_n(r_i,f)Y_n^m(\theta_i,\phi_i)
\end{equation}
is the boundary-aware modal transfer function from the $(n,m)$-th SH component to the $i$-th microphone, and $n_i(t,f)$ denotes measurement noise. By stacking the microphone signals into $\mathbf{x}(t,f)\in\mathbb{C}^{M\times 1}$ and the Ambisonic coefficients into
\begin{equation}
\mathbf{a}_N(t,f)
=
[a_{00}(t,f),a_{1,-1}(t,f),a_{1,0}(t,f),a_{1,1}(t,f),\ldots,a_{N,N}(t,f)]^T
\in\mathbb{C}^{(N+1)^2\times 1},
\end{equation}
the observation model can be written compactly as
\begin{equation}
\mathbf{x}(t,f)
=
\mathbf{A}(\Omega,f)\mathbf{a}_N(t,f)
+
\mathbf{n}(t,f),
\label{eq:forward_model}
\end{equation}
where $\mathbf{A}(\Omega,f)\in\mathbb{C}^{M\times(N+1)^2}$ is the boundary-aware forward sampling matrix. Its entries are formed by the modal transfers $b_{i,nm}(f)$. Therefore, $\mathbf{A}(\Omega,f)$ is a geometry- and boundary-dependent discretization of the continuous SH-domain sound-field model.

Ambisonic encoding aims to recover the target SH coefficients from microphone observations using an analysis operator:
\begin{equation}
\hat{\mathbf{a}}_N(t,f)
=
\mathbf{W}_N(f)\mathbf{x}(t,f),
\label{eq:encoding}
\end{equation}
where $\mathbf{W}_N(f)\in\mathbb{C}^{(N+1)^2\times M}$ is the frequency-dependent encoding matrix. In contrast, $\mathbf{A}(\Omega,f)$ maps SH coefficients to microphone-domain observations and can be regarded as a synthesis operator. Under an ideal fully sampled configuration, where $\mathbf{A}(\Omega,f)$ is known accurately and has full column rank, a linear encoder may be chosen as a left inverse satisfying
\begin{equation}
\mathbf{W}_N(f)\mathbf{A}(\Omega,f)
=
\mathbf{I}_{(N+1)^2},
\label{eq:inverse}
\end{equation}
where $\mathbf{I}_{(N+1)^2}$ is the identity matrix. A conventional choice is the Moore--Penrose pseudo-inverse,
\begin{equation}
\mathbf{W}_N(f)
\approx
\mathbf{A}^{\dagger}(\Omega,f).
\label{eq:pseudoinverse_encoder}
\end{equation}
Substituting Eq.~\eqref{eq:pseudoinverse_encoder} and Eq.~\eqref{eq:forward_model} into Eq.~\eqref{eq:encoding} gives
\begin{equation}
\hat{\mathbf{a}}_N(t,f)
=
\mathbf{A}^{\dagger}(\Omega,f)\mathbf{A}(\Omega,f)\mathbf{a}_N(t,f)
+
\mathbf{A}^{\dagger}(\Omega,f)\mathbf{n}(t,f).
\label{eq:pseudoinverse_reconstruction}
\end{equation}
Equation~\eqref{eq:pseudoinverse_reconstruction} reveals two sources of error: the deviation of $\mathbf{A}^{\dagger}\mathbf{A}$ from the identity operator, and the amplification of measurement noise by $\mathbf{A}^{\dagger}$.

This pseudo-inverse formulation is particularly problematic for sparse arrays. First, compact devices often do not satisfy $M\geq (N+1)^2$, which is necessary for $\mathbf{A}(\Omega,f)$ to have full column rank. In this case, $\mathbf{A}(\Omega,f)$ has a non-trivial right null space $\mathcal{N}(\mathbf{A})$. For any $\mathbf{z}\in\mathcal{N}(\mathbf{A})$, the coefficient vectors $\mathbf{a}_N$ and $\mathbf{a}_N+\mathbf{z}$ produce identical microphone observations in the absence of noise, and the inverse mapping is non-unique. Second, even when a local pseudo-inverse exists, the singular values of $\mathbf{A}(\Omega,f)$ can become small because of irregular sensor placement or zeros and notches in the boundary-dependent radial responses. Let
$\mathbf{A}=\mathbf{U}\mathbf{\Sigma}\mathbf{V}^H$ denote the singular value decomposition of $\mathbf{A}$. The noise contribution to the pseudo-inverse estimate is then
\begin{equation}
\mathbf{A}^{\dagger}\mathbf{n}
=
\mathbf{V}\mathbf{\Sigma}^{\dagger}\mathbf{U}^H\mathbf{n},
\label{eq:svd_noise}
\end{equation}
where small singular values $\sigma_\ell(f)$ lead to amplification factors $1/\sigma_\ell(f)$. For a full-column-rank matrix, the sensitivity of the inverse is characterized by
\begin{equation}
\kappa(\mathbf{A}(\Omega,f))
=
\frac{\sigma_{\max}(\mathbf{A}(\Omega,f))}
{\sigma_{\min}(\mathbf{A}(\Omega,f))}.
\label{eq:condition_number}
\end{equation}
This condition number becomes unbounded when the matrix is rank deficient. Consequently, boundary effects, sparse spatial sampling, and irregular microphone placement yield a strongly frequency-dependent inverse problem: low-frequency higher-order modes can be weakly observed because of small radial terms, whereas high-frequency modes may be spatially aliased or insufficiently constrained by the available microphone density.

To avoid direct inverse filtering, this work reformulates sparse microphone-to-HOA encoding as a geometry-conditioned statistical estimation problem. Here, \emph{variable-topology arrays} refer to arrays with a fixed number of microphones $M$ but different microphone coordinates $\Omega$. Fixing $M$ preserves a consistent network input dimension, while allowing the microphone positions to vary across sparse and irregular three-dimensional configurations within a prescribed physical aperture.

\begin{figure}
\centering
\includegraphics[width=\textwidth]{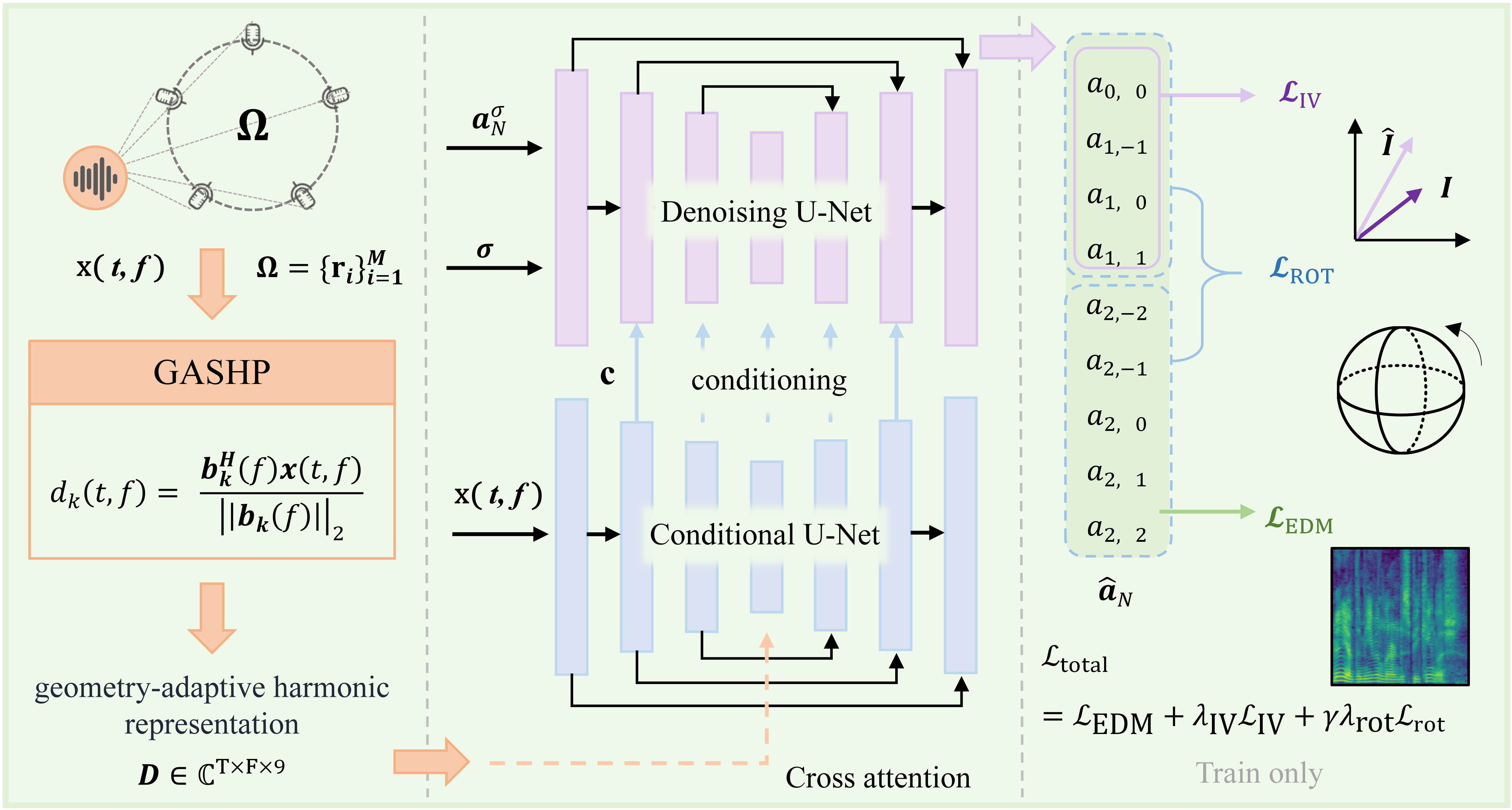}
\caption{Overview of the proposed DiffM2A framework. During training, the target Ambisonic coefficients $\mathbf{a}_N$ are corrupted with Gaussian noise to obtain $\mathbf{a}_N^\sigma$. During inference, the coefficient estimate is initialized from Gaussian noise and iteratively denoised using DPM-Solver.}
\label{fig:framework}
\end{figure}

\section{Proposed method}
\label{sec:method}

As illustrated in Fig. \ref{fig:framework}, DiffM2A estimates Ambisonic coefficients from sparse multi-channel observations through a model-informed geometry-adaptive front-end and a conditional diffusion backbone. The front-end, termed Geometry-Adaptive Spherical Harmonic Projection (GASHP), uses analytical boundary models to project array-dependent microphone observations into a fixed-size, SH-indexed feature representation. This representation reduces the variability induced by microphone geometry before being processed by a dual-branch Elucidated Diffusion Model (EDM). The diffusion backbone is used as a conditional estimator for the underdetermined microphone-to-Ambisonic mapping, while the spatial constraints retain phase-sensitive and rotation-consistent Ambisonic structures. The formulation is written for an arbitrary target order $N$; in the experiments, it is instantiated for FOA and SOA, where five-microphone inputs lead to underdetermined inverse problems.

\subsection{Geometry-Adaptive Spherical Harmonic Projection (GASHP)}
\label{subsec:GASHP}

The preliminary analysis shows that directly inverting the SH-domain forward matrix is problematic for sparse arrays. GASHP is therefore designed as a front-end representation rather than as a closed-form encoder. Its role is to convert the array-dependent microphone observation $\mathbf{x}(t,f)\in\mathbb{C}^{M\times 1}$ into a fixed-size modal tensor that is aligned with the target Ambisonic order and still carries the geometry and boundary information needed by the downstream estimator. Here, $\Omega=\{\mathbf{r}_i\}_{i=1}^{M}$ denotes the sensor coordinates, with $\mathbf{r}_i=(r_i,\theta_i,\phi_i)$ expressed in the array-centered spherical coordinate system. The channel count $M$ is fixed for a given network instance, while the coordinate layout $\Omega$ may vary across arrays.

Following the SH-domain forward model in Sec.~\ref{sec:preliminaries}, the target order is truncated at $N$, giving $K=(N+1)^2$ ACN-ordered modal components. For each frequency bin, GASHP first builds an analytical steering matrix $\mathbf{B}(\Omega,f)=[\mathbf{b}_1(f),\dots,\mathbf{b}_K(f)]\in\mathbb{C}^{M\times K}$,
where the $i$-th entry of the $k$-th column is the boundary-dependent modal transfer $b_{i,nm}(f)=R_n(r_i,f)Y_n^m(\theta_i,\phi_i)$ associated with the corresponding SH mode $k\leftrightarrow(n,m)$. Thus, $\mathbf{B}(\Omega,f)$ is the truncated, implementation-level counterpart of the forward matrix $\mathbf{A}(\Omega,f)$ in Eq.~\eqref{eq:forward_model}. In simulated open-array experiments, $R_n(r_i,f)$ follows Eq.~\eqref{eq:radial_open}; for rigid-sphere recordings, it follows Eq.~\eqref{eq:radial_rigid}. This construction makes the front-end explicitly dependent on the chosen boundary model, while leaving the downstream network architecture unchanged.

Instead of computing $\mathbf{B}^{\dagger}$, GASHP applies an energy-normalized matched-filter projection. For each mode, the microphone observation is correlated with the corresponding analytical steering vector:
\begin{equation}
    d_k(t,f) =  \frac{\mathbf{b}_k^H(f) \mathbf{x}(t,f)}{\lVert \mathbf{b}_k(f) \rVert_2}, \quad k \in \{1, \dots, K\},
\label{eq:gashp_projection}
\end{equation}
where $(\cdot)^H$ denotes the conjugate transpose and $\|\cdot\|_2$ denotes the $\ell_2$-norm. The projected modal vector is
\begin{equation}
    \mathbf{d}(t,f) = [d_1(t,f), \dots, d_K(t,f)]^T,
\end{equation}
and the full GASHP feature tensor is obtained by stacking it over time and frequency:
\begin{equation}
    \mathbf{D}=\{\mathbf{d}(t,f)\}_{t,f}\in\mathbb{C}^{T\times F\times K}.
\end{equation}
Because all arrays are projected onto the same $K$ modal channels, $\mathbf{D}$ provides a common representation for variable array layouts. 

The projection should not be interpreted as an unbiased Ambisonic estimate. To make this explicit, write the microphone observation within the truncated SH subspace as
\begin{equation}
    \mathbf{x}(t,f) = \sum_{k'=1}^K \mathbf{b}_{k'}(f) a_{k'}(t,f) + \mathbf{n}(t,f).
\end{equation}
Substituting this model into Eq.~\eqref{eq:gashp_projection} gives
\begin{align}
    d_k(t,f) &= \frac{\mathbf{b}_k^H(f) \mathbf{x}(t,f)}{\lVert \mathbf{b}_k(f) \rVert_2} \notag \\
    &= \sum_{k'=1}^K \underbrace{\left( \frac{\mathbf{b}_k^H(f) \mathbf{b}_{k'}(f)}{\lVert \mathbf{b}_k(f) \rVert_2} \right)}_{\Gamma_{k, k'}(f)} a_{k'}(t,f) + \underbrace{\frac{\mathbf{b}_k^H(f) \mathbf{n}(t,f)}{\lVert \mathbf{b}_k(f) \rVert_2}}_{\eta_k(t,f)},
\label{eq:gashp_derivation}
\end{align}
or, in compact form,
\begin{equation}
    \mathbf{d}(t,f) = \mathbf{\Gamma}(f) \mathbf{a}_N(t,f) + \mathbf{\eta}(t,f),
\label{eq:coupled_matrix}
\end{equation}
where $\mathbf{\Gamma}(f)\in\mathbb{C}^{K\times K}$ is the frequency-dependent modal coupling matrix induced by sparse and non-orthogonal sampling, and $\mathbf{\eta}(t,f)\in\mathbb{C}^{K\times 1}$ is the projected composite noise. Since Eq.~\eqref{eq:gashp_projection} normalizes by $\lVert\mathbf{b}_k(f)\rVert_2$ rather than $\lVert\mathbf{b}_k(f)\rVert_2^2$, the diagonal entries of $\mathbf{\Gamma}(f)$ are not constrained to be one. GASHP therefore produces a coupled modal observation, not a unit-gain modal reconstruction. The remaining scaling error, cross-mode leakage, and boundary mismatch are intentionally left to the conditional diffusion estimator in Sec.~\ref{subsec:diffusion_backbone}.

The vector $\mathbf{\eta}(t,f)$ contains both spatially filtered sensor noise and residual modeling errors caused by SH truncation, boundary mismatch, and sensor uncertainty. For the sensor noise component, if $\mathbf{n}(t,f)$ is zero-mean spatially white noise with covariance $\mathbb{E}[\mathbf{n}(t,f)\mathbf{n}^H(t,f)] = \sigma^2 \mathbf{I}_M$, the variance after normalized mode matching is
\begin{equation}
\mathbb{E}\left[\left|\frac{\mathbf{b}_k^H(f)\mathbf{n}(t,f)}{\lVert \mathbf{b}_k(f) \rVert_2}\right|^2\right] = \frac{\mathbf{b}_k^H(f) \mathbb{E}[\mathbf{n}(t,f)\mathbf{n}^H(t,f)] \mathbf{b}_k(f)}{\lVert \mathbf{b}_k(f) \rVert_2^2} = \frac{\sigma^2 \left(\mathbf{b}_k^H(f)\mathbf{b}_k(f)\right)}{\lVert \mathbf{b}_k(f) \rVert_2^2} = \sigma^2.
\end{equation}
Thus, under this noise model, the projection does not introduce the singular-value-dependent noise amplification associated with $\mathbf{A}^{\dagger}$. This property is the main reason for using GASHP as a front-end: it supplies a boundary-aware and geometry-aligned modal representation while avoiding the most unstable part of pseudo-inverse encoding. The resulting $\mathbf{D}$ is then used together with the raw microphone spectra as the conditioning information for the diffusion denoiser.

\subsection{Dual-Branch Conditional Diffusion Denoiser}
\label{subsec:diffusion_backbone}

As shown in Eq.~\eqref{eq:coupled_matrix}, GASHP produces a coupled modal observation rather than a direct estimate of the Ambisonic coefficients. The remaining modal scaling, cross-mode leakage, and boundary mismatch make an algebraic inversion of $\mathbf{\Gamma}(f)$ undesirable. DiffM2A therefore uses a conditional diffusion denoiser to estimate $\mathbf{a}_N$ from the information jointly provided by the raw microphone spectra and the GASHP representation.

The term ``dual-branch'' refers to the separation between a conditioning branch and a denoising branch. The conditioning branch fuses microphone-domain and GASHP-domain observations, whereas the denoising branch performs EDM-based coefficient estimation under this condition. For the neural representation, microphone observations are stacked as $\mathbf{x}\in\mathbb{C}^{T\times F\times M}$, GASHP features as $\mathbf{D}\in\mathbb{C}^{T\times F\times K}$, and target coefficients as $\mathbf{a}_N\in\mathbb{C}^{T\times F\times K}$, where $K=(N+1)^2$. Complex-valued tensors are implemented by concatenating their real and imaginary parts along the channel dimension. The conditioning branch, implemented as a Conditional U-Net, computes a spatio-spectral feature tensor $\mathbf{c}\in\mathbb{R}^{T\times F\times C_c}$, where $C_c$ denotes the number of conditioning channels:
\begin{equation}
    \mathbf{c} = \mathcal{F}_{\text{cond}}(\mathbf{x}, \mathbf{D}).
\end{equation}
In this design, $\mathbf{x}$ preserves sensor-domain cues that may be attenuated by modal projection, while $\mathbf{D}$ provides a boundary-aware harmonic representation with reduced geometry dependence. The conditioning tensor $\mathbf{c}$ therefore connects the physical forward model, the GASHP front-end, and the statistical estimator.

Conditioned on $\mathbf{c}$, the denoising branch learns a noise-conditional mapping in the Ambisonic coefficient domain. During training, the target $\mathbf{a}_N$ is perturbed by a continuous Gaussian noise schedule:
\begin{equation}
    \mathbf{a}_N^\sigma = \mathbf{a}_N + \sigma\mathbf{\epsilon}, \quad \mathbf{\epsilon} \sim \mathcal{N}(\mathbf{0}, \mathbf{I}),
\end{equation}
where $\sigma$ denotes the diffusion noise level and $\mathbf{\epsilon}$ is a standard Gaussian perturbation in the Ambisonic coefficient domain. The conditional denoiser $D_\theta(\cdot)$, parameterized by network weights $\theta$, predicts the corresponding target coefficient vector from $\mathbf{a}_N^\sigma$, $\sigma$, and $\mathbf{c}$. To improve numerical stability across noise scales, we use the EDM preconditioning parameterization \cite{10.5555/3600270.3602196}:
\begin{equation}
    D_\theta(\mathbf{a}_N^\sigma, \sigma, \mathbf{c}) =
    c_{\text{skip}}(\sigma)\mathbf{a}_N^\sigma
    + c_{\text{out}}(\sigma)
    \mathcal{F}_{\text{den}}\left(c_{\text{in}}(\sigma)\mathbf{a}_N^\sigma, c_{\text{noise}}(\sigma), \mathbf{c}\right),
\label{eq:edm_denoiser}
\end{equation}
where $\mathcal{F}_{\text{den}}$ is the U-Net denoising network. The EDM preconditioning coefficients are defined as
\begin{equation}
\begin{aligned}
&c_{\text{skip}}(\sigma) = \frac{\sigma_{\text{data}}^2}{\sigma^2+\sigma_{\text{data}}^2},
&c&_{\text{out}}(\sigma) = \frac{\sigma\sigma_{\text{data}}}{\sqrt{\sigma^2+\sigma_{\text{data}}^2}},
\\
&c_{\text{in}}(\sigma) = \frac{1}{\sqrt{\sigma^2+\sigma_{\text{data}}^2}},
&c&_{\text{noise}}(\sigma) = \frac{1}{4}\log\sigma,
\\
&\lambda(\sigma) = \frac{\sigma^2+\sigma_{\text{data}}^2}{(\sigma\sigma_{\text{data}})^2}.
\end{aligned}
\label{eq:edm_preconditioning}
\end{equation}
Here, $\sigma_{\text{data}}$ denotes the standard deviation of the normalized target Ambisonic coefficient distribution. The coefficient $c_{\text{in}}$ scales the noisy input to the denoising network, $c_{\text{noise}}$ provides the scalar noise-level embedding, $c_{\text{skip}}$ controls the residual skip path, and $c_{\text{out}}$ scales the predicted residual. The weighting term $\lambda(\sigma)=1/c_{\text{out}}^2(\sigma)$ balances the denoising errors across noise levels. The network parameters $\theta$ are trained using the weighted denoising objective:
\begin{equation}
    \mathcal{L}_{\text{EDM}} = \mathbb{E}_{\mathbf{a}_N, \mathbf{x}, \mathbf{D}, \sigma, \mathbf{\epsilon}}\left[ \lambda(\sigma) \left\lVert D_\theta(\mathbf{a}_N^\sigma, \sigma, \mathbf{c}) - \mathbf{a}_N \right\rVert_2^2 \right].
\end{equation}
Although $\mathbf{D}$ is deterministically computed from $\mathbf{x}$, $\Omega$, and the selected boundary model, it is included in the expectation to make the conditioning variables explicit. In the implementation, the complex Ambisonic coefficients are represented by their real and imaginary components, so the denoising objective is evaluated over the corresponding real-valued channel tensor.

During inference, the conditioning branch first computes $\mathbf{c}$ from the observed microphone spectra and the corresponding GASHP tensor. The Ambisonic coefficients are then generated through iterative reverse denoising steps initialized from Gaussian noise. The backbone therefore serves as a statistical continuation of the preceding physical model.

\subsection{Multi-Tiered Spatial Constraints}
\label{subsec:spatial_constraints}

While the diffusion objective is to reconstruct target HOA coefficients, it does not explicitly encode phase-dependent inter-channel relationships or the transformation structure of Ambisonic representations. This may lead to spatial inconsistency when the microphone observations are sparse. We therefore introduce two complementary spatial regularization tiers: a low-order tier based on active pseudo-intensity cues from the FOA subspace and a higher-order tier based on the rotation behavior of SH coefficient subspaces.

First, the low-order tier uses an active pseudo-intensity vector (IV) loss to enforce phase-sensitive directional energy consistency. For any target order $N\geq 1$, the FOA subset of $\mathbf{a}_N$ provides the pressure and directional components used to compute this cue. The active IV describes the net acoustic energy flow between the zero-order sound pressure $W(t,f)$ and the first-order directional particle-velocity-related vector $\mathbf{v}(t,f) = [Y(t,f), Z(t,f), X(t,f)]^T$:
\begin{equation}
    \mathbf{I}(t,f) = \mathfrak{R}\left\{ W^*(t,f)\mathbf{v}(t,f) \right\},
\label{eq:iv_definition}
\end{equation}
where $\mathfrak{R}\{\cdot\}$ extracts the real part and $(\cdot)^*$ denotes the complex conjugate. Normalizing $\mathbf{I}(t,f)$ by the total energy density $E(t,f) = |W(t,f)|^2 + \lVert \mathbf{v}(t,f) \rVert_2^2$ gives the normalized energy vector $\mathbf{r}_E(t,f) = \mathbf{I}(t,f) / (E(t,f) + \epsilon)$, where $\epsilon > 0$ is a small stabilization constant. The low-order loss penalizes the $\ell_1$ distance between the estimated and reference intensity-related quantities:
\begin{equation}
    \mathcal{L}_{\text{IV}} = \mathbb{E}_{t,f} \left[ \left\lVert \hat{\mathbf{I}}(t,f) - \mathbf{I}(t,f) \right\rVert_1 + \lambda_{ev} \left\lVert \hat{\mathbf{r}}_E(t,f) - \mathbf{r}_E(t,f) \right\rVert_1 \right],
\label{eq:l_iv}
\end{equation}
where $\hat{\mathbf{I}}(t,f)$ and $\hat{\mathbf{r}}_E(t,f)$ are evaluated directly from the estimated HOA coefficient vector $\hat{\mathbf{a}}_N(t,f)$.

Second, the higher-order tier uses a rotational consistency loss motivated by the transformation properties of HOA coefficients under the $\mathrm{SO}(3)$ group. When an acoustic scene undergoes a global 3D rotation $\mathbf{R} \in \mathrm{SO}(3)$, the ideal HOA coefficients transform linearly within each independent $(2n+1)$-dimensional subspace via real-valued Wigner-D matrices \cite{osti_249653}. During rotation-augmented training, the source positions and room geometry are rotated consistently, and the corresponding microphone spectra and GASHP tensor are used as the rotated network input. To enforce equivariant behavior, we minimize the discrepancy between the network prediction for the rotated input and the Wigner-D transformed prediction for the original input within each $n$-th order subspace ($n \in \{1, \dots, N\}$):
\begin{equation}
    \mathcal{L}_{\text{rot}} = \sum_{n=1}^N \mathbb{E}_{t,f} \left\lVert \hat{\mathbf{a}}_n^{\text{rot}}(t,f) - \mathbf{D}_n(\mathbf{R})\hat{\mathbf{a}}_n(t,f) \right\rVert_1,
\label{eq:l_rot}
\end{equation}
where $\mathbf{D}_n(\mathbf{R})$ represents the $(2n+1)$-dimensional real-valued Wigner-D matrix block. The transformation is applied to the real-valued channel representation of the complex STFT coefficients, equivalently to the real and imaginary parts using the same Wigner-D block.

The final training objective combines the diffusion loss with the two spatial regularizers:
\begin{equation}
    \mathcal{L}_{\text{total}} = \mathcal{L}_{\text{EDM}} + \lambda_{\text{IV}}\mathcal{L}_{\text{IV}} + \gamma \lambda_{\text{rot}}\mathcal{L}_{\text{rot}},
\label{eq:l_total}
\end{equation}
where $\lambda_{\text{IV}}$ and $\lambda_{\text{rot}}$ are balancing hyperparameters, and $\gamma \in \{0,1\}$ activates the rotational term only for batches with rotation augmentation.

Together, these two constraints provide complementary regularization without requiring DOA metadata or explicit source priors. The intensity loss encourages consistency of low-order directional energy and phase-sensitive inter-channel relationships, whereas the rotational loss encourages the estimated HOA coefficients to follow the expected $\mathrm{SO}(3)$ transformation behavior across harmonic subspaces. In combination, these constraints help reduce order-wise inconsistency and encourage more structured Ambisonic estimates under sparse sampling.

\section{Experiments}
\label{sec:exp}

\subsection{Datasets}
\noindent \textbf{(a) Simulated dataset.}~We use analytical room impulse response (RIR) simulations as the primary training and controlled evaluation data, since they provide access to target Ambisonic signals and allow explicit control over array geometry, source layout, and reverberation. To evaluate geometry generalization across unseen configurations, randomized acoustic scenes are generated using the HARP method \cite{saini2025harp}. Shoebox rooms are sampled with dimensions between $4 \times 4 \times 4$~m and $12 \times 12 \times 8$~m, yielding reverberation times ($T_{60}$) from 0.05~s to 0.9~s. The training set contains 5000 acoustic scenes augmented with 100 distinct MAs. For validation and testing, we generate 100 acoustic scenes augmented with 50 MAs each. The MAs used for training, validation, and testing are disjoint, so that the evaluation arrays represent unseen geometric topologies. In all configurations, five microphones are randomly positioned within a sphere of radius $0.09$~m, with inter-microphone distances constrained to $[0.02, 0.18]$~m. Single- and dual-source mixtures are synthesized by convolving simulated RIRs with clean speech from the WSJ0 corpus \cite{garofolo1993csr}; the source-to-array distance is constrained to be at least $2.0$~m, and dual-source scenes use an azimuthal separation of at least $25^\circ$. During training, rotation augmentation is applied to 50\% of the batches ($\gamma=1$). With the coordinate system centered at the MA origin, the source positions and room geometry are globally rotated by a random 3D rotation $\mathbf{R} \in \mathrm{SO}(3)$, while the local array coordinate frame is kept fixed. The corresponding Wigner-D matrix blocks $\mathbf{D}_n(\mathbf{R})$ are computed using the Ivanic-Ruedenberg recurrence algorithm \cite{osti_249653}.

\vspace{\baselineskip}

\noindent \textbf{(b) LOCATA dataset.}~
To assess performance under measured acoustic conditions, we use multichannel recordings from the LOCATA challenge dataset \cite{8448644}. The recordings were captured in an enclosed acoustic laboratory ($7.1 \times 9.8 \times 3$~m, $T_{60}\approx0.55$~s) with moderate reverberation and multipath propagation. All recordings are downsampled to 16~kHz. A sparse five-microphone input is formed by selecting five sensors from the 32-channel rigid-sphere Eigenmike array with radius $a=0.042$~m. The five selected sensors are used only as the sparse input, while the full 32-channel Eigenmike recording is used to generate the reference SOA signals offline using the standard Eigenmike encoding matrix with radial equalization filters. This reference is therefore an array-encoded Ambisonic reference rather than an ideal continuous-field ground truth (GT). Because the Eigenmike is mounted on a rigid spherical baffle, the measured sound field includes rigid-sphere scattering and diffraction effects. Accordingly, the GASHP front-end uses the rigid-sphere radial response in Eq.~\eqref{eq:radial_rigid} rather than the free-field response used for simulated open-array training, while the downstream neural architecture is kept unchanged.

\subsection{Evaluation metrics}
We use objective metrics to evaluate signal fidelity, spectral accuracy, spatial correlation, and binaural spatial cues. Temporal fidelity is measured by the Scale-Invariant Signal-to-Distortion Ratio (SI-SDR). Let $s_k[l] = \text{iSTFT}\{a_k(t, f)\}$ and $\hat{s}_k[l] = \text{iSTFT}\{\hat{a}_k(t, f)\}$ denote the discrete time-domain signals of the $k$-th channel ($k \in \{1, \dots, K\}$). Stacking $L$ time-domain samples into channel vectors $\bm{s}_k = [s_k[1], \dots, s_k[L]]^T \in \mathbb{R}^{L \times 1}$ and $\hat{\bm{s}}_k \in \mathbb{R}^{L \times 1}$, the overall SI-SDR is computed as
\begin{equation}
    \text{SI-SDR} = \frac{1}{K} \sum_{k=1}^{K} 10 \log_{10} \frac{\Vert{}\alpha_k \bm{s}_k\Vert{}_2^2}{\Vert{}\alpha_k \bm{s}_k - \hat{\bm{s}}_k\Vert{}_2^2},
\end{equation}
where $\alpha_k = \frac{\hat{\bm{s}}_k^T \bm{s}_k}{\Vert{}\bm{s}_k\Vert{}_2^2}$ is the optimal scaling factor.

Within the Ambisonic domain, spectral accuracy and spatial correlation are assessed using Magnitude Spectrum Error (Mag.Err) and channel-wise Coherence (Coh). The Mag.Err is computed in decibels (dB) across $K$ Ambisonic channels, $T$ time frames, and $F$ frequency bins:
\begin{equation}
    \text{Mag.Err} = \frac{1}{K \cdot T \cdot F} \sum_{k=1}^K \sum_{t=1}^T \sum_{f=1}^F \left| 20\log_{10}|a_k(t,f)| - 20\log_{10}|\hat{a}_k(t,f)| \right|,
\end{equation}
where $\lvert{}\cdot\rvert{}$ denotes the absolute value, and $a_k(t,f)$ and $\hat{a}_k(t,f)$ denote the reference and estimated complex spectra of the $k$-th Ambisonic channel, respectively. Spatial correlation and phase alignment are assessed using channel-wise Coherence (Coh):
\begin{equation}
    \text{Coh} = \frac{1}{K \cdot F} \sum_{k=1}^K \sum_{f=1}^F \frac{\left| \sum_{t=1}^T a_k(t,f) \hat{a}_k^*(t,f) \right|^2}{\left(\sum_{t=1}^T |a_k(t,f)|^2\right) \left(\sum_{t=1}^T |\hat{a}_k(t,f)|^2\right)},
\end{equation}
where $*$ denotes the complex conjugate operator.

To evaluate binaural spatial cues, the synthesized Ambisonic signals are decoded into binaural representations using Head-Related Transfer Functions (HRTFs) \cite{article2017}. Let $x_L(t,f)$ and $x_R(t,f)$ denote the left- and right-ear complex spectra of the decoded binaural signals, respectively. The interaural level difference (ILD) in dB is defined as
\begin{equation}
    \text{ILD}(t,f) = 10 \log_{10} \frac{|x_L(t,f)|^2}{|x_R(t,f)|^2}.
\end{equation}
We compute the root mean square error (RMSE) between the estimated and reference ILD maps, denoted by ILD.Err:
\begin{equation}
    \text{ILD.Err} = \sqrt{\frac{1}{T \cdot F} \sum_{t=1}^T \sum_{f=1}^F \left( \text{ILD}(t,f) - \widehat{\text{ILD}}(t,f) \right)^2}.
\end{equation}

\subsection{Implementation details}
\label{subsec:details}
All audio signals are resampled to 16~kHz and converted into complex STFT representations, which are cropped or padded to $256 \times 256$ time-frequency bins. The conditioning branch receives the real and imaginary parts of the five-channel microphone observations as a 10-channel input and incorporates the GASHP representation through the conditioning pathway described in Sec.~\ref{subsec:diffusion_backbone}. The denoising branch predicts the real and imaginary parts of the target Ambisonic coefficients, corresponding to 8 output channels for FOA and 18 output channels for SOA. The generative backbone is a 2D U-Net with a base channel dimension of 32 and channel multipliers of $\{1, 2, 4, 8\}$. Each scale contains one ResNet block, and 4-head self-attention is applied at the three lowest spatial resolutions. Models are trained for 200 epochs using the Adam optimizer with a learning rate of $1 \times 10^{-4}$ and momentum parameters $(\beta_1,\beta_2)=(0.9,0.99)$. The diffusion process follows the Karras noise schedule \cite{10.5555/3600270.3602196}, with $\sigma_{\text{min}} = 0.002$ and $\sigma_{\text{max}} = 80.0$. The data variance scaling factor $\sigma_{\text{data}}$ is fixed at $0.2$. During training, the continuous noise level $\sigma$ is stochastically drawn from a log-normal distribution with hyperparameters $\mu_{\text{noise}} = -1.0$ and $\sigma_{\text{noise}} = 1.2$. The loss weights are set to $\lambda_{ev}=0.01$, $\lambda_{\text{IV}}=0.2$, and $\lambda_{\text{rot}}=0.5$. During inference, samples are generated using a 64-step DPM-Solver. All experimental results are reported as the average of five independent runs.

\noindent \textbf{Real-World Domain Adaptation:}
For LOCATA evaluation, the GASHP front-end adopts the rigid-sphere boundary formulation in Eq.~\eqref{eq:radial_rigid}, with the physical radius fixed at $a=0.042$~m. This setting changes only the physics-informed front-end; the network architecture and target Ambisonic format remain identical to those used for simulated open-array experiments. To account for the domain gap between analytical simulation and measured rigid-sphere recordings, the downstream neural networks are fine-tuned on the LOCATA training partition for 30 epochs with a learning rate of $1\times10^{-5}$. An exponential moving average (EMA) of the network parameters is maintained during fine-tuning with a decay factor of $\beta_{\text{EMA}}=0.999$, and all reported LOCATA results are obtained using the EMA weights. For fair comparison, all DL-based baselines are adapted with the same LOCATA partition and fine-tuning protocol under their respective training objectives.

\section{Results}
\subsection{Comparison with baselines}

We compare DiffM2A with four representative baselines: Parametric, AmbiSpatial, Gen-A, and Attention-based models \cite{9795678,10890048,10887869,beyond}. The parametric baseline serves as a reference for spatial filtering when directional estimates are available. The other three baselines represent recent DL-based encoding approaches. All neural baselines use the same sparse input channels, STFT representation, FOA/SOA targets, and train/validation/test splits as DiffM2A. For baseline methods originally designed for FOA or array-dependent settings, we adapt their output heads to match the target dimensionality and retrain them under the same supervision.

\begin{table}[ht!]
\centering
\caption{Objective evaluation results on the simulated dataset for single- and dual-source scenarios.}
\label{tab:main_results}
\begin{tabular*}{\textwidth}{@{\extracolsep{\fill}} ll cccc cccc @{}}
\toprule
\multirow{3}{*}{\textbf{Task}} & \multirow{3}{*}{\textbf{Method}} & \multicolumn{4}{c}{\textbf{Single Source}} & \multicolumn{4}{c}{\textbf{Dual Source}} \\
\cmidrule(lr){3-6} \cmidrule(lr){7-10}
& & \textbf{SI-SDR} & \textbf{Coh} & \textbf{Mag.\ Err} & \textbf{ILD.Err} & \textbf{SI-SDR} & \textbf{Coh} & \textbf{Mag.\ Err} & \textbf{ILD.Err} \\
& & (dB) $\uparrow$ & $\uparrow$ & (dB) $\downarrow$ & (dB) $\downarrow$ & (dB) $\uparrow$ & $\uparrow$ & (dB) $\downarrow$ & (dB) $\downarrow$ \\
\midrule
\multirow{5}{*}{\textbf{FOA}} 
& Param.           & 7.82 & \textbf{0.695} & \textbf{5.93} & \textbf{3.29} & 4.24 & \underline{0.630} & \underline{8.85} & \underline{4.50} \\
& AmbiSpatial      & 10.57 & 0.629 & 8.99 & 4.91 & 6.83 & 0.551 & 12.32 & 5.78 \\
& Gen-A           & 11.09 & 0.649 & 8.81 & 4.74 & 7.59 & 0.566 & 11.57 & 5.59 \\
& Attention-based  & \underline{12.43} & 0.654 & 7.90 & 3.82 & \underline{8.76} & 0.593 & 10.23 & 4.97 \\
& \textbf{Proposed} & \textbf{15.42} & \underline{0.687} & \underline{6.42} & \underline{3.57} & \textbf{11.59} & \textbf{0.650} & \textbf{8.21} & \textbf{4.28} \\
\midrule
\multirow{5}{*}{\textbf{SOA}} 
& Param.           & 5.96 & \underline{0.458} & \underline{7.87} & \underline{4.10} & 1.21 & \underline{0.301} & \underline{10.98} & \underline{5.51} \\
& AmbiSpatial      & 8.61 & 0.366 & 8.93 & 5.82 & 4.47 & 0.197 & 14.41 & 6.80 \\
& Gen-A           & 9.18 & 0.379 & 8.79 & 5.58 & 6.01 & 0.223 & 13.50 & 6.38 \\
& Attention-based  & \underline{10.64} & 0.311 & 8.47 & 4.89 & \underline{7.46} & 0.254 & 12.66 & 5.92 \\
& \textbf{Proposed} & \textbf{13.51} & \textbf{0.461} & \textbf{7.60} & \textbf{3.98} & \textbf{9.74} & \textbf{0.405} & \textbf{9.27} & \textbf{5.02} \\
\bottomrule
\end{tabular*}
\end{table}

\begin{figure}
\centering
\includegraphics[width=\textwidth]{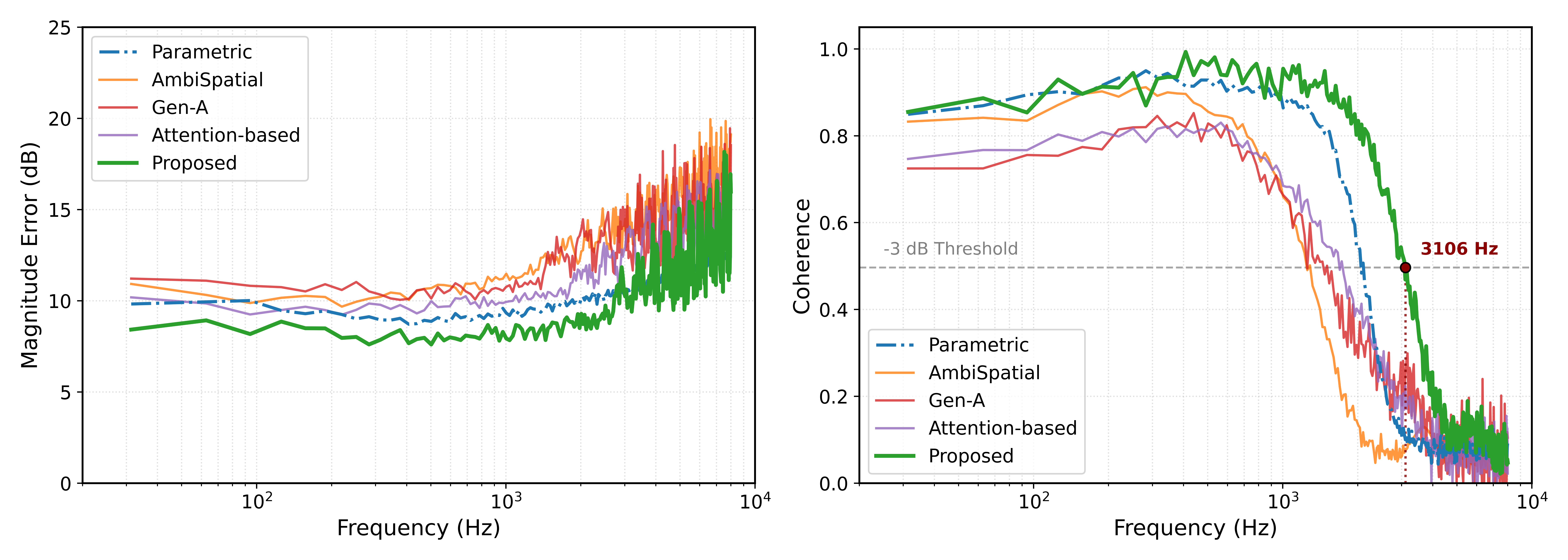}
\caption{Comparison of mean magnitude spectrum error and coherence metrics under the simulated dual-source $\text{SOA}$ configuration.}
\label{fig:mag_coh_analysis}
\end{figure}

\begin{table}[!]
\centering
\caption{Objective evaluation results on the LOCATA dataset for single- and dual-source scenarios.}
\label{tab:locata_results}
\begin{tabular*}{\textwidth}{@{\extracolsep{\fill}} ll cccc cccc @{}}
\toprule
\multirow{3}{*}{\textbf{Task}} & \multirow{3}{*}{\textbf{Method}} & \multicolumn{4}{c}{\textbf{Single Source}} & \multicolumn{4}{c}{\textbf{Dual Source}} \\
\cmidrule(lr){3-6} \cmidrule(lr){7-10}
& & \textbf{SI-SDR} & \textbf{Coh} & \textbf{Mag.\ Err} & \textbf{ILD.Err} & \textbf{SI-SDR} & \textbf{Coh} & \textbf{Mag.\ Err} & \textbf{ILD.Err} \\
& & (dB) $\uparrow$ & $\uparrow$ & (dB) $\downarrow$ & (dB) $\downarrow$ & (dB) $\uparrow$ & $\uparrow$ & (dB) $\downarrow$ & (dB) $\downarrow$ \\
\midrule
\multirow{5}{*}{\textbf{FOA}}
& Param.           & 0.85 & \underline{0.71} & 5.82 & 3.88 & -0.81 & \underline{0.60} & 7.15 & 4.62 \\
& AmbiSpatial      & 3.82 & 0.58 & 4.34 & 3.31 & 1.98 & 0.47 & 5.58 & 4.12 \\
& Gen-A            & 4.65 & 0.63 & 3.92 & 2.95 & 2.85 & 0.52 & 5.12 & 3.61 \\
& Attention-based  & \underline{5.21} & 0.68 & \underline{3.65} & \underline{2.68} & \underline{3.42} & 0.58 & \underline{4.68} & \underline{3.24} \\
& \textbf{Proposed} & \textbf{8.45} & \textbf{0.84} & \textbf{2.32} & \textbf{1.52} & \textbf{6.88} & \textbf{0.79} & \textbf{2.91} & \textbf{1.95} \\ 
\midrule
\multirow{5}{*}{\textbf{SOA}}
& Param.           & -0.42 & 0.50 & 6.95 & 4.35 & -2.15 & 0.41 & 8.32 & 5.42 \\
& AmbiSpatial      & 2.45 & 0.46 & 5.23 & 3.92 & 0.72 & 0.36 & 6.48 & 4.88 \\
& Gen-A            & 3.12 & 0.51 & 4.85 & 3.48 & 1.48 & 0.42 & 5.95 & 4.35 \\
& Attention-based  & \underline{3.88} & \underline{0.57} & \underline{4.42} & \underline{3.12} & \underline{2.11} & \underline{0.48} & \underline{5.42} & \underline{3.98} \\
& \textbf{Proposed} & \textbf{7.23} & \textbf{0.76} & \textbf{2.88} & \textbf{1.88} & \textbf{5.62} & \textbf{0.71} & \textbf{3.35} & \textbf{2.48} \\ 
\bottomrule
\end{tabular*}
\end{table}

Table \ref{tab:main_results} summarizes the objective evaluation results for both single- and dual-source scenarios. Although the parametric method achieves high coherence in FOA configurations, its SI-SDR is significantly lower across all conditions. This indicates that linear spatial filters cannot effectively isolate the direct path from complex multipath reflections in reverberant environments, leading to severe temporal smearing and low-frequency noise amplification. While DL baselines mitigate these distortions, the proposed DiffM2A consistently outperforms them, achieving the best balance between spectral fidelity and spatial accuracy.


Fig. \ref{fig:mag_coh_analysis} shows the frequency-domain performance for the dual-source SOA configuration. Due to spatial aliasing in sparse arrays, the spatial coherence of DL baselines drops around 800--1000~Hz, accompanied by comb-filtering artifacts in magnitude error. To quantify the operational range, we evaluate the effective upper frequency limit using the standard $-3\text{ dB}$ coherence threshold. Under this metric, the AmbiSpatial, Gen-A, and Attention-based baselines drop below the threshold at $1317\text{ Hz}$, $1631\text{ Hz}$, and $1945\text{ Hz}$, respectively, while the Parametric method sustains coherence up to $2133\text{ Hz}$. DiffM2A extends this range to $3106\text{ Hz}$ while maintaining a lower magnitude error baseline. This result suggests that the GASHP front-end and spatial regularization reduce high-frequency aliasing and phase ambiguity, although the coherence still approaches a noise floor above 5~kHz.

Table~\ref{tab:locata_results} presents the objective performance on the real-world LOCATA dataset. Due to measurement noise, room effects, and rigid-baffle scattering, all neural methods degrade compared with the simulation results. DiffM2A nevertheless maintains higher SI-SDR and coherence than the DL baselines, reaching 5.62~dB SI-SDR and 0.71 coherence under the dual-source SOA configuration.

The parametric baseline shows a larger performance drop on LOCATA than in the controlled simulations. A likely reason is model mismatch: real recordings contain array scattering, calibration errors, early reflections, and spatially correlated reverberation that are not fully captured by idealized directional or diffuse-field assumptions. These factors can increase the sensitivity of spatial filtering to covariance estimation and matrix inversion errors. In contrast, DiffM2A learns a conditional mapping from boundary-aware projected observations, which may reduce its reliance on these idealized assumptions.

To validate sound-field consistency, we project the estimated SOA coefficients back to the 5-channel microphone space. As shown in Table~\ref{tab:reprojection_errors}, DiffM2A achieves the highest reprojection SI-SDR (8.05~dB) and the lowest spectral magnitude error (3.14~dB), outperforming all baseline methods and even exceeding the GT SOA reference (6.86~dB). This result indicates that the second-order HOA representation is limited by truncation errors when representing microphone signals, especially over a high-frequency range. In contrast, DiffM2A leverages learned generative priors to resolve the ambiguity of the underdetermined inverse mapping and recover physically plausible HOA coefficients that better match the microphone observations. Consequently, the proposed framework establishes a more consistent mapping between sparse microphone measurements and spatial harmonic representations.

\begin{table}[t]
\centering
\caption{Comparison of sensor-domain reprojection errors under the dual-source SOA configuration.}
\label{tab:reprojection_errors}
\begin{tabular}{lcc}
\toprule
\textbf{Method} & \textbf{SI-SDR (dB)} $\uparrow$ & \textbf{Mag.Err (dB)} $\downarrow$ \\
\midrule
Ground Truth & 6.86 & 3.71 \\
\midrule
Param. & 1.01 & 7.84 \\
AmbiSpatial & 3.94 & 5.43 \\
Gen-A & 4.71 & 4.97 \\
Attention-based & 5.48 & 4.53 \\
\textbf{Proposed} & \textbf{8.05} & \textbf{3.14} \\
\bottomrule
\end{tabular}
\end{table}

\begin{table}[!htp]
\centering
\caption{Ablation study for dual-source SOA encoding.}
\label{tab:ablation_loss}
\resizebox{0.7\textwidth}{!}{
\begin{tabular}{l cccc}
\toprule
\textbf{Configuration} & \textbf{SI-SDR} (dB) $\uparrow$ & \textbf{Coh} $\uparrow$ & \textbf{Mag.\ Err} (dB) $\downarrow$ & \textbf{ILD.Err} (dB) $\downarrow$ \\
\midrule
$\mathcal{L}_{\text{EDM}}$                                                             & 6.75  & 0.224 & 11.52 & 6.08 \\
$\mathcal{L}_{\text{EDM}} + \mathcal{L}_{\text{IV}}$                              & 7.32  & 0.288 & 10.64 & 5.62\\
$\mathcal{L}_{\text{EDM}} + \mathcal{L}_{\text{rot}}$                                 & 8.54  & 0.332 & 10.08 & 5.38 \\
$\mathcal{L}_{\text{EDM}} + \mathcal{L}_{\text{IV}} + \mathcal{L}_{\text{rot}}$ & \textbf{9.74}  & \textbf{0.405} & \textbf{9.27}  & \textbf{5.02} \\ \hline
w/o GASHP  & 7.16  & 0.272 & 10.92 & 5.75 \\
GASHP + U-Net  & 7.21  & 0.282 & 10.97 & 5.87 \\
\bottomrule
\end{tabular}}
\end{table}

\subsection{Ablation study and array generalization analysis}

We evaluate the contribution of each module under the simulated dual-source SOA configuration, as summarized in Table \ref{tab:ablation_loss}. The baseline ($\mathcal{L}_{\text{EDM}}$) recovers the main spectral envelopes but gives lower coherence and larger ILD errors. Adding the low-order physical anchor ($\mathcal{L}_{\text{EDM}} + \mathcal{L}_{\text{IV}}$) improves SI-SDR and coherence by regularizing the active pseudo-intensity flow. Adding the geometric anchor ($\mathcal{L}_{\text{EDM}} + \mathcal{L}_{\text{rot}}$) further improves the metrics by encouraging rotation-consistent behavior across harmonic subspaces. Removing the sound-field projection front-end (w/o GASHP) degrades all metrics, which supports the role of GASHP in reducing array-dependent variability. The full configuration gives the best overall performance.

To evaluate the necessity of the generative backbone, we introduce a deterministic baseline, GASHP + U-Net. This variant adopts an architecture matched to the conditioning branch and directly regresses the target HOA coefficients from the GASHP features $\mathbf{D}$. It uses an MSE loss with the same multi-tiered spatial constraints ($\mathcal{L}_{\text{IV}}$ and $\mathcal{L}_{\text{rot}}$) as DiffM2A, but omits the diffusion process. As shown in Table~\ref{tab:ablation_loss}, this model yields a modest 7.21 dB SI-SDR and 0.282 spatial coherence. The proposed generative framework outperforms it by 2.53 dB and 0.123, respectively, confirming that probabilistic diffusion modeling is crucial for overcoming spatial ambiguity and recovering finer spatial-spectral details under severe undersampling.

\begin{figure}
\centering
\includegraphics[width=\textwidth]{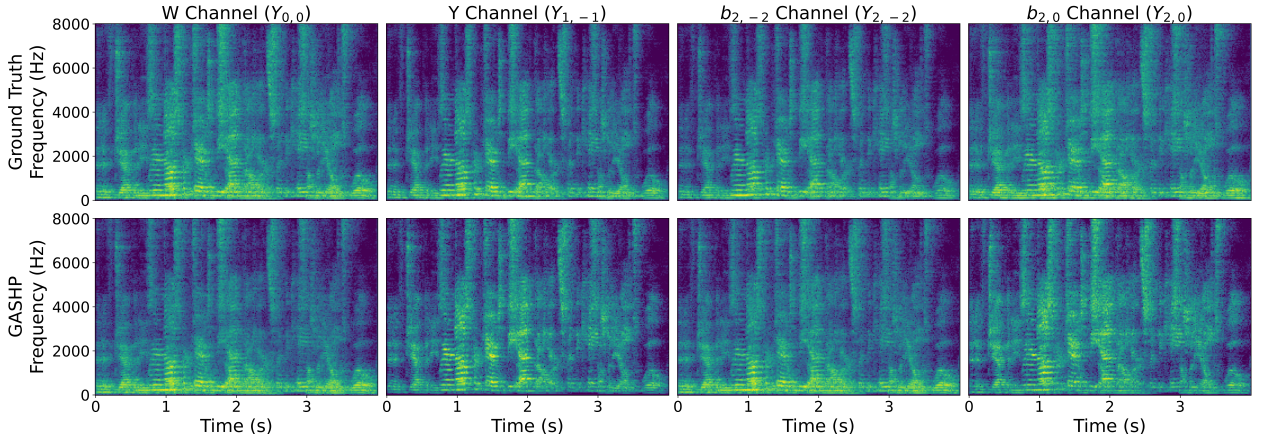}
\caption{Comparison of STFT spectrograms of the ground-truth HOA coefficients (top) and the corresponding GASHP-projected modal representations (bottom) for four representative harmonic channels ($W$, $Y$, $b_{2,-2}$, and $b_{2,0}$) under the simulated dual-source configuration. }
\label{fig:gashp_analysis}
\end{figure}

To illustrate the behavior of the geometry-adaptive modal projection, we visualize the modal representations produced by GASHP and compare them with the corresponding reference Ambisonic coefficients. As illustrated in Fig. \ref{fig:gashp_analysis}, GASHP preserves the dominant spectro-temporal structures while transforming irregular microphone observations into a common modal representation. Compared with the reference coefficients, whose harmonic channels exhibit mode-dependent structures, the projected modal features are more homogeneous across channels. This behavior is consistent with Eq.~\eqref{eq:coupled_matrix}, where the frequency-dependent modal coupling matrix $\mathbf{\Gamma}(f)$ introduces residual cross-mode leakage under sparse spatial sampling. GASHP therefore reduces array-dependent variability, while the downstream diffusion model handles the remaining modal decoupling and spatial deblurring.

\begin{table}[!htp]
\centering
\caption{Comparison on unseen arrays in SI-SDR (dB).}
\label{tab:diffm2a_comparison}
\resizebox{0.65\textwidth}{!}{
\begin{tabular}{lccccc|c}
\hline
\multicolumn{1}{l}{\textbf{Method}} & \textbf{C-5-6} & \textbf{C-5-4} & \textbf{L-5-4} & \textbf{L-5-3} & \textbf{Avg.} & \textbf{Params (M)}  \\ \hline
Param.  & 2.42  & 3.26  & 1.35  & 1.51  & 2.14  & --        \\
AmbiSpatial        & 4.53  & 5.13  & 4.40  & 4.51  & 4.64  & 8.12        \\
Gen-A             & 5.48  & 6.05  & 5.21  & 5.34  & 5.52  & 9.10     \\
Attention-based  & 7.54 & 7.73  & 7.15  & 7.23  & 7.41  & 1.19    \\ \hline

w/o GASHP                             & 7.66  & 7.75  & 7.16  & 7.31  & 7.47  & 8.53   \\
\textbf{Proposed}                    & \textbf{9.81} & \textbf{10.10} & \textbf{9.54} & \textbf{9.65} & \textbf{9.78} & 8.66 \\ \hline
\end{tabular}}
\end{table}

Furthermore, we evaluate the cross-geometry generalization of DiffM2A on a set of unseen conventional MAs, as summarized in Table \ref{tab:diffm2a_comparison}, where C-5-6 represents a circular array with 5 channels and a radius of 6~cm, and L-5-4 denotes a linear array with 5 channels and 4~cm spacing. DiffM2A achieves the highest average SI-SDR of 9.78~dB. Removing the sound-field projection front-end (w/o GASHP) reduces the average SI-SDR to 7.47~dB, indicating that the geometry-normalized modal representation helps mitigate array-domain shifts. The conventional attention-based model is compact, but its reliance on accurate acoustic metadata may limit its robustness when geometry information is uncertain.

\begin{figure}
\centering
\includegraphics[width=\textwidth]{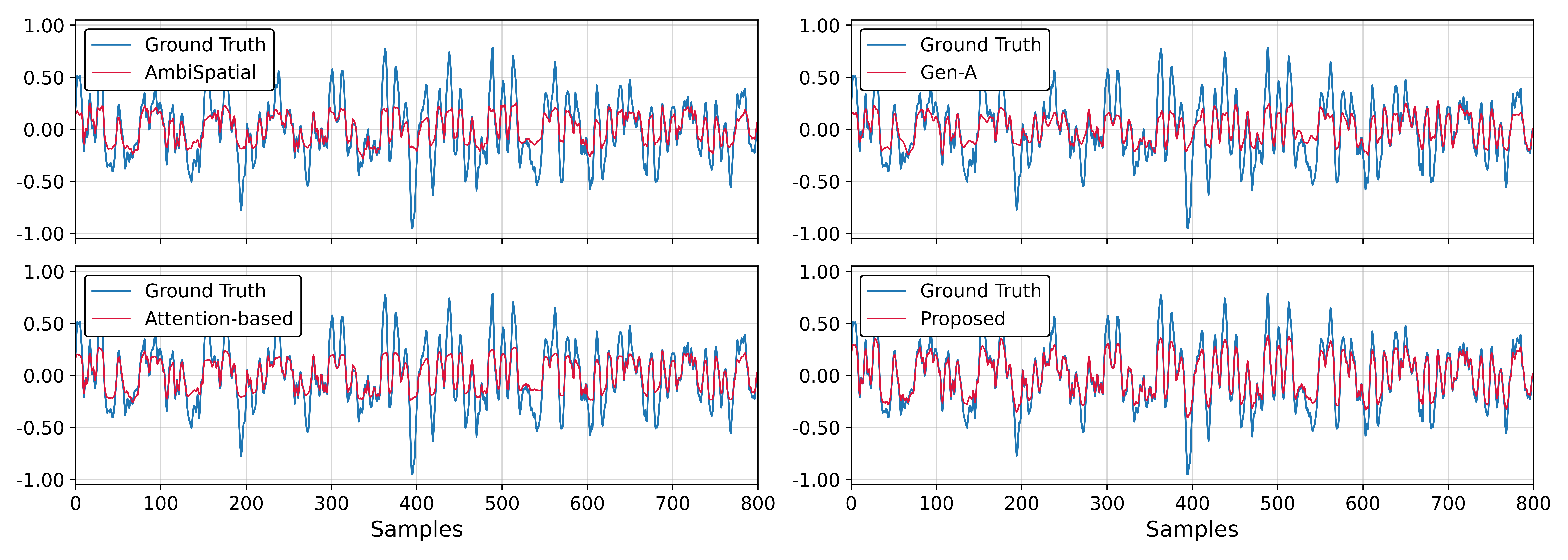}
\caption{Time-domain waveforms of the reconstructed $W$-channel under the simulated dual-source $\text{SOA}$ configuration.}
\label{fig:time_domain}
\end{figure}

\begin{figure}
\centering
\includegraphics[width=\textwidth]{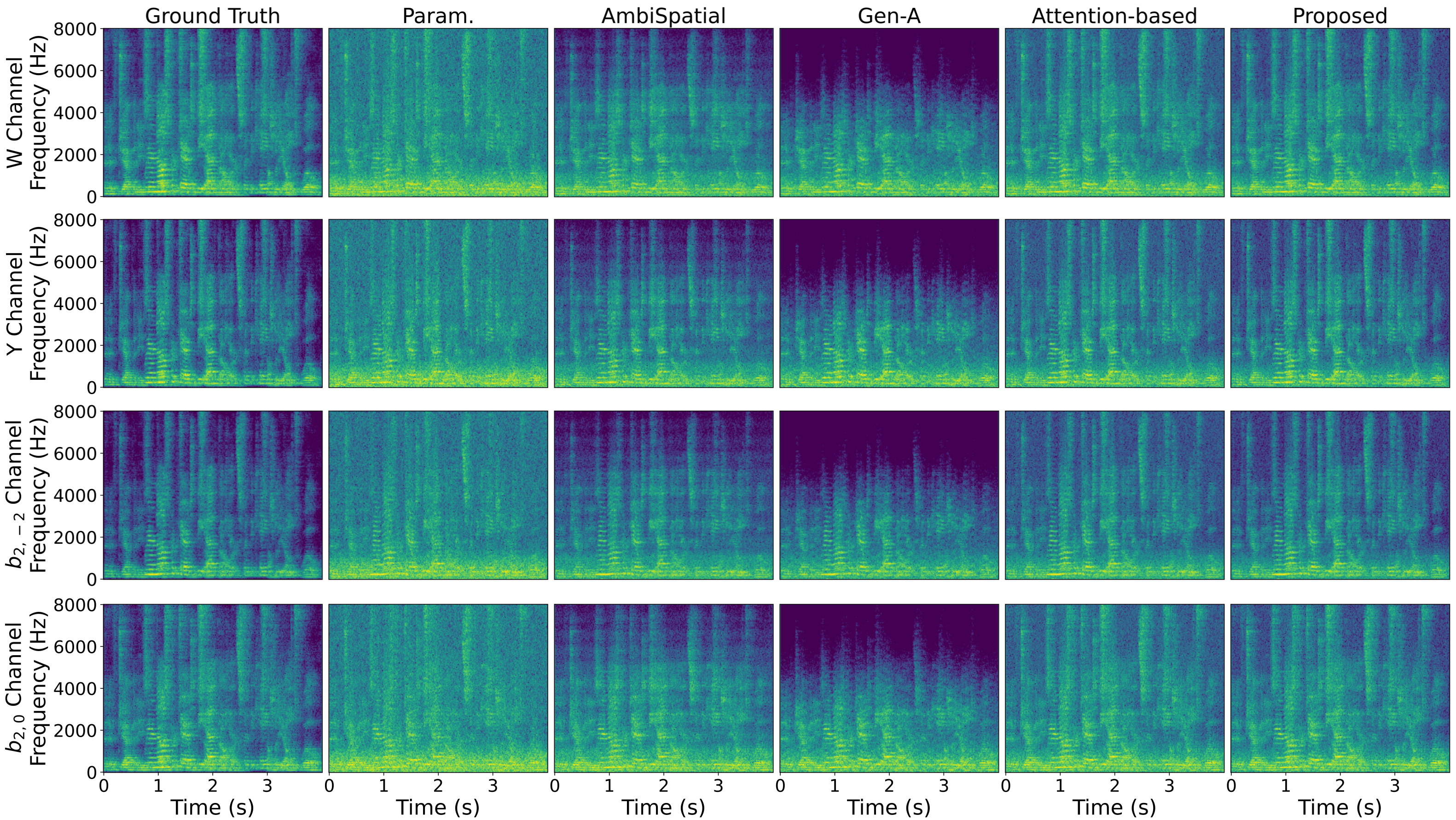}
\caption{Comparison of STFT spectrograms for reconstructed Ambisonic channels ($W$, $Y$, $b_{2,-2}$, and $b_{2,0}$) across different methods under the simulated dual-source configuration.}
\label{fig:across_analysis}
\end{figure}

\subsection{Analysis across Ambisonic orders}

To provide a more intuitive comparison of the spatial encoding quality across different methods, Fig. \ref{fig:time_domain} presents the $W$-channel time-domain waveforms reconstructed via inverse STFT (iSTFT) from various encoded Ambisonic signals under the dual-source SOA configuration.

As observed, AmbiSpatial and Gen-A capture the general temporal periodicity but show amplitude compression and underestimated high-energy transient peaks, which is consistent with over-smoothing in deterministic regression. The attention-based model shows similar tracking limitations at finer signal scales. In contrast, DiffM2A follows the target waveform more closely in this example, with better preservation of high-amplitude peaks and localized structural fluctuations. This behavior is consistent with reduced conditional-mean smoothing and may benefit temporal-envelope preservation in downstream spatial audio rendering.

To examine cross-order structural preservation, Fig.~\ref{fig:across_analysis} shows the STFT spectrograms of reconstructed Ambisonic channels, including low-order FOA components ($W$ and $Y$) and second-order SOA components ($b_{2,-2}$ and $b_{2,0}$). Figs.~\ref{fig:gashp_analysis} and \ref{fig:across_analysis} are generated from the same audio sample. The conventional parametric method shows noise-floor amplification and reduced spectral contrast in the mid-to-high frequency range, which is consistent with the sensitivity of plane-wave assumptions to overlapping wavefronts in reverberant environments. Among the DL-based methods, Gen-A shows high-frequency attenuation above $4\text{ kHz}$, especially in the second-order channels. AmbiSpatial and the attention-based model recover the broadband energy envelopes, but their harmonic structures appear over-smoothed. This observation is consistent with the tendency of deterministic networks to produce conditional-mean estimates when sparse microphone observations do not uniquely determine the higher-order components.

In contrast, DiffM2A better preserves harmonic structures and detailed spectral variations across different Ambisonic orders. By formulating the encoding task as a denoising-based generative sampling process conditioned on the GASHP representation $\mathbf{D}$, the proposed model alleviates the spatial ambiguity caused by sparse microphone observations and avoids the over-smoothed solutions commonly produced by deterministic regression models. Furthermore, the spatial regularizers reduce inconsistency among output channels: the intensity-vector loss constrains the low-order directional energy distribution, while the rotation loss enforces the expected $\mathrm{SO}(3)$-equivariant relationship among second-order Ambisonic components.

\begin{table}[htbp]
\centering
\caption{Comparison of the training cost and average inference time (evaluated per 4-second audio sample at 16~kHz). The neural network-based models are evaluated on a single NVIDIA GeForce RTX 5090 GPU, whereas the parametric baseline is executed on an AMD EPYC 9554 64-Core CPU.}
\label{tab:computation_cost}
\begin{tabular}{lcc}
\hline
\textbf{Method} & \textbf{Training Cost} & \textbf{Inference Time} \\ \hline
Parametric  & -- & 0.35 s \\
AmbiSpatial & Medium & 1.23 s \\
Gen-A  & Medium & 1.36 s \\
Attention-based  & Low & 0.93 s \\
\textbf{Proposed} & High & 2.66 s \\ \hline
\end{tabular}
\end{table}

As shown in Table \ref{tab:computation_cost}, the parametric method has the shortest inference time ($0.35\text{ s}$) due to its analytical formulation, but it yields lower reconstruction scores in the preceding evaluations and depends on reliable acoustic metadata. The DL-based baselines require approximately $0.93$--$1.36\text{ s}$ per 4-second sample, whereas DiffM2A requires $2.66\text{ s}$ due to its iterative reverse sampling process. This additional computational cost is the main trade-off for the improved reconstruction quality observed in the previous experiments. For underdetermined conditions, the tested neural methods recover broad spectral envelopes to some extent, but AmbiSpatial and Gen-A show less inter-channel phase detail, and the attention-based model still has limited high-frequency resolution. DiffM2A retains sharper harmonic tracks and more coherent multi-order spectral textures in the visualized examples, at the cost of longer inference time.



\section{Conclusion}

This paper presented DiffM2A, a geometry-adaptive framework for Ambisonic encoding from sparse MAs with variable topology. The GASHP front-end projects microphone observations into a boundary-aware SH modal representation without explicit inverse filtering, and the dual-branch EDM estimates Ambisonic coefficients from both microphone-domain and modal-domain information. The spatial constraints further regularize low-order intensity cues and higher-order rotational structure. Experiments on simulated rooms and LOCATA recordings show improved FOA/SOA encoding performance over the evaluated baselines, together with better generalization across unseen array geometries and mismatched boundary conditions. The remaining limitation is computational cost, which motivates future work on faster diffusion sampling and model distillation.

Future work will investigate faster sampling strategies and model distillation to reduce the inference cost of the diffusion backbone for real-time or edge-device deployment. Another direction is to extend the analytical GASHP front-end with more detailed boundary models, such as head-scattering and torso-diffraction effects, to better match wearable spatial audio devices.

\section*{Data availability}
Data will be made available on request.

\section*{Acknowledgments}
This work was supported by the National Natural Science Foundation of China under Grant Nos. 62271401, 62171373, and 61831019.







\printcredits

\bibliographystyle{cas-model2-names}

\bibliography{cas-refs}



\end{document}